# Instrumental neutron activation analysis of an enriched $^{28}$Si single-crystal

G. D'Agostino(1), L. Bergamaschi(1), L. Giordani(1), G. Mana(2), M. Oddone(3)

(1) Istituto Nazionale di Ricerca Metrologica (INRIM) - Unit of Radiochemistry and Spectroscopy, c/o Department of Chemistry, University of Pavia, via Taramelli 12, 27100 Pavia, Italy

(2) Istituto Nazionale di Ricerca Metrologica (INRIM), Strada delle Cacce 91, 10135 Torino, Italy.

(3) Department of Chemistry – Radiochemistry Area, University of Pavia, via Taramelli 12, 27100 Pavia, Italy

Corresponding author: G. D'Agostino, g.dagostino@inrim.it, telephone +39 0382 526252, fax +39 0382 423578

**Abstract:** The determination of the Avogadro constant plays a key role in the redefinition of the kilogram in terms of a fundamental constant. The present experiment makes use of a silicon single-crystal highly enriched in $^{28}$Si that must have a total impurity mass fraction smaller than a few parts in $10^9$. To verify this requirement, we previously developed a relative analytical method based on neutron activation for the elemental characterization of a sample of the precursor natural silicon crystal WASO 04. The method is now extended to fifty-nine elements and applied to a monoisotopic $^{28}$Si single-crystal that was grown to test the achievable enrichment. Since this crystal was likely contaminated, this measurement tested also the detection capabilities of the analysis. The results quantified contaminations by Ge, Ga, As, Tm, Lu, Ta, W and Ir and, for a number of the detectable elements, demonstrated that we can already reach the targeted 1 ng/g detection limit.

## Introduction

The kilogram is the last base unit of the Système International still defined in relation to an artifact. As the proposed new definition will be based on the Planck constant [1, 2], $h$, the determination of the Avogadro constant, $N_A$, is of paramount importance to obtain a $h$ value via the molar Planck constant, $N_A h$, which is very accurately known [3]. The present method to derive $N_A$ is based on counting the Si atoms in a crystal by measuring its density, isotopic composition and unit cell volume of a silicon crystal [4]. In particular, the latest and most accurate $N_A$ measurement relies on the availability of nearly perfect and highly pure silicon crystals very much enriched in $^{28}$Si. The relative uncertainty of the result is evaluated to be about $3 \times 10^{-8}$ [5].

Imperfections such as point-like defects due to contaminant atoms strain the crystal and change its mass. Therefore, experimental data concerning the contaminants are required to endorse the purity of the Avogadro silicon crystals. Accordingly, it is essential to verify that the total impurity mass fraction is below a few parts in $10^9$ or, in case of a higher value, to quantify the total impurity with an uncertainty smaller than the above limit.

Activation analysis is one of the classical tools to analyze silicon. In particular, Neutron Activation Analysis (NAA) and Activation Analysis with Charged Particle (CPAA) are remarkably suited for the determination of contaminants in silicon. Low detection limits can be reached for a large number of elements, including those having a low atomic number, e.g. B, C, N and F [6].
Since CPAA is certainly more demanding in terms of equipment compared to NAA, the latter became the most applied method for the characterization of silicon. Moreover, thanks to the small and short-lived activity produced in silicon by the neutron irradiation, very low detection limits can be reached instrumentally, i.e. without the radiochemical separation. For this reason, Instrumental Neutron Activation Analysis (INAA) is frequently preferred to Radiochemical Neutron Activation Analysis (RNAA). In fact, several methods based on INAA have been developed and applied, mainly for the determination of impurities in silicon materials produced by semiconductor industries [7-11]. For instance, the INAA of silicon wafers performed by Kim *et alii* achieved detection limits ranging between 2 fg/g and 0.3 ng/g for forty-nine elements [7]. The irradiation lasted 72 h and the thermal neutron flux was $3.7 \times 10^{13}$ $cm^{-2}$ $s^{-1}$, with a thermal to fast flux ratio of about 20.

Taking into account that the detection limits reported in literature fulfill our target, we developed a INAA method to check the contamination of the silicon crystals used for the $N_A$ determination. The method was preliminary tested on a sample of the Avogadro natural silicon crystal WASO 04. The investigation concerned twenty-nine elements and reached detection limits ranging between 1 pg/g and 10 μg/g [12].

Since the potential contaminant elements are eighty-nine, i.e. the naturally occurring elements excluding the silicon, we aimed at increasing the number of measured elements. Thus, starting from the method applied to the WASO 04, we extended the analysis to fifty-nine elements. Given the uniqueness, value, and reduced availability of the remnants of the $^{28}Si$ material used for the $N_A$ determination, we decided to further check the detection capabilities of the extended analysis on a $^{28}Si$ crystal which was grown only to verify the achievable enrichment and supposed to be contaminated. The present paper focuses on the results of this measurement.

## The sample

Fig. 1 shows a picture of the sample of the enriched $^{28}Si$ single-crystal we used for testing the neutron activation method. Diameter, length and mass are 13 mm, 42 mm and 11.6 g, respectively.

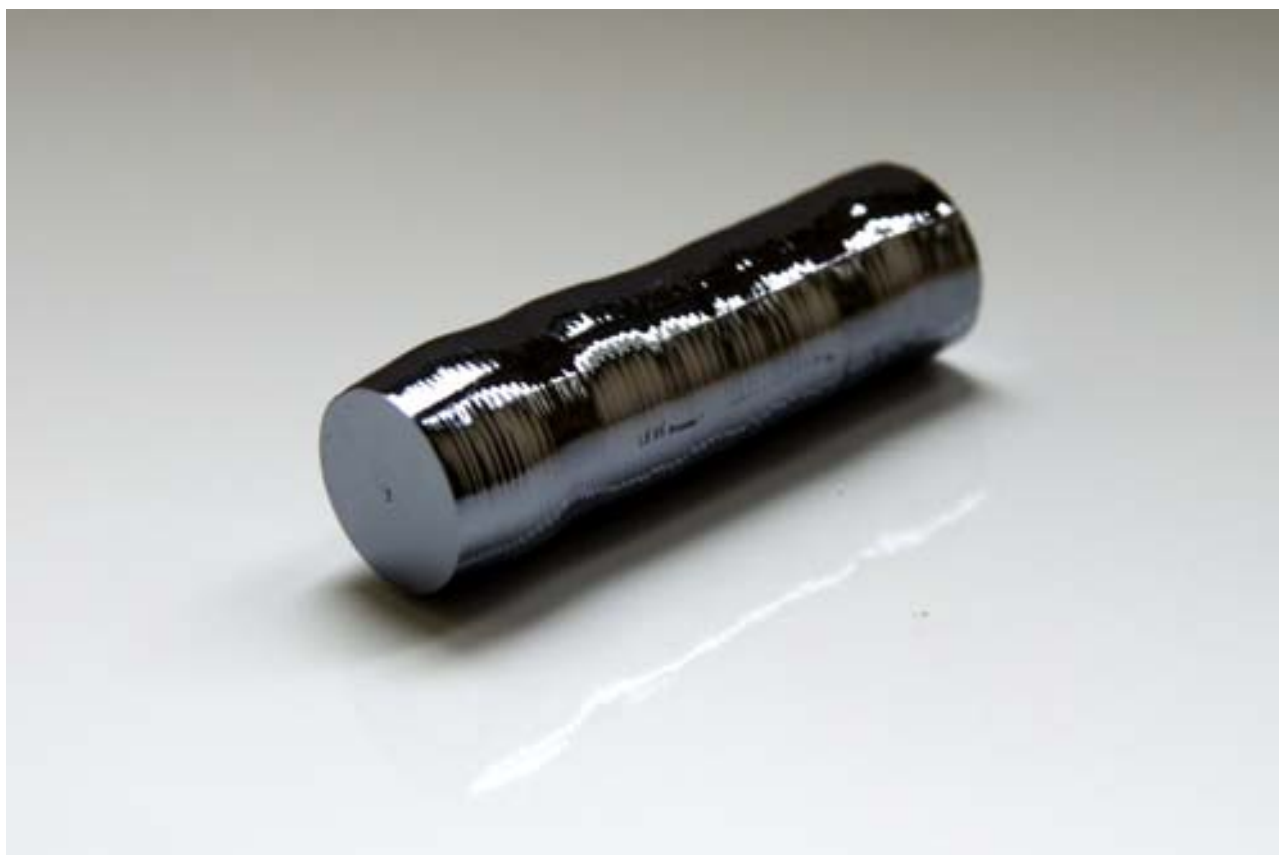

**Fig. 1** The sample of the enriched $^{28}Si$

This sample (Si28-21Pr10.2, part 2) was taken from a $^{28}$Si single-crystal grown in 2012 by the Leibniz-Institut für Kristallzüchtung (IKZ, Berlin, Germany). The isotopic enrichment of the $SiF_4$ gas was carried out by the ElectroChemical Plant (ECP, Zelenogorsk, Russia). The conversion of the $SiF_4$ gas to silane, its chemical purification, and the poly-crystal deposition was carried out by the Institute of Chemistry of High-Purity Substances (IChHPS, Nizhniy Novgorod, Russia). The Physikalisch Technische Bundesanstalt (PTB, Braunschweig, Germany) measured the isotopic composition of the poly-crystal by the isotope dilution mass spectrometry. The results confirmed an enrichment in $^{28}$Si higher than 99.985%. In addition, the amount of oxygen and carbon within the single-crystal was determined by infrared spectrometry at PTB. The concentrations were found to be about $2 \times 10^{17}$ cm$^{-3}$ and $3 \times 10^{15}$ cm$^{-3}$, respectively. The methods used for the measurement of the isotopic composition and for the determination of the impurities are described in [13, 14] and [15], respectively.

## Application of the method

A detailed description of the neutron activation method used for the analysis can be found in [12]. Here, for convenience of the reader, we recall the measurement equation, i.e.

$$\frac{N_{\text{sam}}}{N_{\text{st}}} = \frac{\left.\dfrac{n_{\text{c}}}{e^{\lambda t_{\text{d}}}(1-e^{\lambda t_{\text{c}}})}\dfrac{1}{\varepsilon}\dfrac{t_{\text{c}}}{t_{\text{live}}}\right|_{\text{sam}}}{\left.\dfrac{n_{\text{c}}}{e^{\lambda t_{\text{d}}}(1-e^{\lambda t_{\text{c}}})}\dfrac{1}{\varepsilon}\dfrac{t_{\text{c}}}{t_{\text{live}}}\right|_{\text{st}}}, \tag{1}$$

where subscripts sam and st indicate the sample and standard, respectively. In addition, $N$ is the number of the target nuclei of the contaminant isotope, $\varepsilon$ and $n_{\text{c}}$ are the fraction of the emitted gamma photons and the net counts stored by the multichannel analyzer in the full-energy peak, $t_{\text{d}}$, $t_{\text{c}}$ and $t_{\text{live}}$ are the decay, counting and live times and $\lambda = \ln(2)/t_{1/2}$ is the decay constant, where $t_{1/2}$ is the half-life time of the gamma emitting radionuclide.

The main neutron capture reactions in silicon crystals are $^{30}$Si(n, γ)$^{31}$Si and $^{29}$Si(n, p)$^{29}$Al. The half-lives of $^{31}$Si and $^{29}$Al are about 2.6 h and 6.6 min, respectively. Extremely small amount of $^{24}$Na is produced by $^{28}$Si through the neutron capture reaction $^{28}$Si(n, αp)$^{24}$Na. The half-life of $^{24}$Na, is 15 h. It is worth to notice that the gamma emission due to the matrix production of $^{30}$Si and $^{29}$Si becomes negligible when the silicon sample is highly enriched in $^{28}$Si.

Traceable solutions of pure substances were used to prepare seventeen standards. The first ten, ML1 ÷ ML10, were prepared for the detection of medium-lived radionuclides, while the remaining seven, LL1 ÷ LL7 were prepared for the detection of long-lived radionuclides. Each standard consists of weighted amount of multi-element solutions which are pipetted onto filter papers rolled up as a cylinder and inserted in polyethylene vials. Before sealing the vials, the pipetted solutions were evaporated to dryness using an infrared lamp in a fume hood under ambient conditions. The multi-element solutions were home-made by adding certified single-element solutions to suitable mass fractions. In particular, the amount of each element was adjusted to have a good counting statistics with the standards located at about 8 cm far from the head of the detector during gamma spectrometric measurements. The resultant amounts of the standard elements resulted to be orders of magnitude higher than the corresponding impurity elements which are known to be present in the filter papers and in the polyethylene vials.

The silicon sample was sealed in a polyethylene vial and inserted in an aluminum container together with ML2, ML3, ML4, ML5, ML6, ML7 and ML10. The remaining ten standards were inserted in a different aluminum container. The containers were sent to the irradiation facility for a neutron bombardment lasting 6 hours. The irradiation was carried at the central thimble of the 250 kW TRIGA Mark II reactor at the Laboratory of Applied Nuclear Energy (LENA) of the University of Pavia. The thermal neutron flux was about $6 \times 10^{12}$ cm$^{-2}$ s$^{-1}$ and the thermal to fast flux ratio was about 11. After the irradiation, the containers were left to cool until the activity decayed to safe values.

Before performing the gamma spectrometric measurements, the vials containing the standards were rinsed with diluted nitric acid and deionized water in order to remove the possible contamination due to handling. Since the goal was to measure only the bulk contamination, the silicon sample was removed from its irradiation vial and etched to eliminate the possible contamination of the external layers. Accordingly, the silicon sample was washed with trichloroethylene, acetone and deionized water, etched for 25 min with a solution 10:1 of nitric acid (assay 67-69%) and hydrofluoric acid (assay 47-51%), and finally rinsed in deionized water, ethylalcohol and acetone. The loss of mass was about 140 mg, corresponding to the removal of a surface layer having a thickness of about 40 μm. Lastly, the silicon sample was sealed in a non-irradiated vial for the gamma counting.

The counting facility consists of an automatic system including a sample changer, a coaxial HPGe detector ORTEC® GEM50P4-83 (relative efficiency 50%, resolution 1.90 keV FWHM at 1332 keV), a digital signal processor ORTEC® DSPEC jr 2.0, and a personal computer running the software for data acquisition and processing ORTEC® Gamma Vision 6.0. The energy, resolution and efficiency calibrations were carried out by a standard multi-gamma source in two counting positions, in contact with (position 0) and 8 cm far (position 8) from the head of the detector.

The gamma spectra were sequentially measured in three different periods. The first data record was collected after 3 days from the end of the irradiation and concerned with the silicon sample and the standards ML1 ÷ ML10; the counting times were 3 h for the silicon and 8 h for the standards at the counting positions 0 and 8, respectively. The second data record was collected after 16 days from the end of the irradiation and concerned only with the silicon sample; the counting time was 24 h at the counting position 0. Finally, the last data record was collected after 24 days from the end of the irradiation and concerned with the standards LL1 ÷ LL7; the counting time was 24 h at the position 8.

The number of target nuclei within the standard is $N_{\mathrm{st}} = \theta_{\mathrm{std}}\, m_{\mathrm{std}}\, N_{\mathrm{A}} / \mathrm{M}$, where $\theta_{\mathrm{std}}$ is the isotopic mole fraction, $m_{\mathrm{std}}$ and $M$ are the mass and the atomic mass of the element, and $N_{\mathrm{A}}$ is the Avogadro constant. The number of target nuclei of the same isotope in the silicon sample, $N_{\mathrm{st}}$, is calculated according to (1).

## Results

The contamination of the crystal could originate before, during or after the isotope separation of the $SiF_4$ gas by the centrifuge cascade. In the latter case, i.e. when the contamination occurs after the isotopic separation, the isotopic composition of the contaminant elements is the natural one. On the contrary, i.e. when the contamination occurs before or during the isotopic separation, the contaminant elements have an isotopic composition different from the natural one.

Thus, we report in table 1 the experimental results both in terms of mass fraction of the detected isotope, $w_{\mathrm{iso}}$, and, in case of natural composition, of the relevant element, $w_{\mathrm{ele}}$. The name of the standard, the isotopic mole fraction in the standard, the half-life of the radioactive nucleus, the energy of the detected gamma peak and the neutron capture reaction are also given. The uncertainties include only the component due to the counting statistics and the detection limits have been calculated according to the Currie's method [16, 17].

The greatest effort was devoted to the achievement of the smallest detection limits for the larger number of elements. Therefore, the effect of efficiency, geometry, self-shielding and non-homogeneity of the neutron flux during irradiation and self-absorption during gamma counting could affect the results up to a few tens per cent.

The quantified isotopes are $^{23}$Na, $^{71}$Ga, $^{75}$As, $^{76}$Ge, $^{169}$Tm, $^{176}$Lu, $^{181}$Ta, $^{186}$W and $^{191}$Ir; the relevant mass fractions are $(1.38 \pm 0.02) \times 10^{-9}$, $(9.62 \pm 0.50) \times 10^{-11}$, $(4.75 \pm 0.07) \times 10^{-10}$, $(2.98 \pm 0.08) \times 10^{-7}$, $(1.02 \pm 0.08) \times 10^{-10}$, $(3.95 \pm 0.56) \times 10^{-14}$, $(4.23 \pm 0.18) \times 10^{-12}$, $(8.77 \pm 0.99) \times 10^{-12}$ and $(2.00 \pm 0.33) \times 10^{-12}$, respectively. The values of $^{23}$Na could be overestimated due to possible interferences due to $^{28}$Si via $^{28}$Si(n, αp)$^{24}$Na and $^{76}$Ge via $^{76}$Ge(n,γ)$^{77}$Ge. In the latter case, the gamma emission at 1368.4 keV of $^{77}$Ge interferes with the gamma emission at 1368.7 keV of $^{24}$Na. Moreover, the value of $^{169}$Tm could be overestimated due to possible interference due to $^{181}$Ta via $^{181}$Ta(n,γ)$^{182}$Ta. In particular, the gamma emission at 84.68 keV of $^{182}$Ta interferes with the gamma emission at 84.25 keV of $^{170}$Tm. In addition, the detection limit of $^{197}$Au could be biased due to suspected problems occurred during the preparation of the Au standard.

| Isotope | Standard | $\theta_{std}$ | $t_{1/2}$ | Peak / keV | Reaction | $w_{iso}$ | $w_{ele}$ |
|---|---|---|---|---|---|---|---|
| $^{23}$Na | ML2 | 1.000000 | 15 h | 1368.7 | $^{23}$Na(n,γ)$^{24}$Na [(a)] | $(1.38 \pm 0.02) \times 10^{-9}$ | $(1.38 \pm 0.02) \times 10^{-9}$ |
| $^{41}$K | ML2 | 0.067302 | 12 h | 1524.77 | $^{41}$K(n,γ)$^{42}$K | $\leq 3.3 \times 10^{-10}$ | $\leq 4.9 \times 10^{-9}$ |
| $^{45}$Sc | LL1 | 1.000000 | 84 d | 889.4 | $^{45}$Sc(n,γ)$^{46}$Sc | $\leq 2.6 \times 10^{-12}$ | $\leq 2.6 \times 10^{-12}$ |
| $^{46}$Ca | ML1 | 0.000040 | 4.5 d | 1297.1 | $^{46}$Ca(n,γ)$^{47}$Ca | $\leq 4.1 \times 10^{-11}$ | $\leq 1.0 \times 10^{-6}$ |
| $^{47}$Ti | ML3 | 0.074400 | 3.3 d | 159.4 | $^{47}$Ti(n,p)$^{47}$Sc | $\leq 1.7 \times 10^{-9}$ | $\leq 2.3 \times 10^{-8}$ |
| $^{50}$Cr | LL1 | 0.043450 | 28 d | 320.03 | $^{50}$Cr(n,γ)$^{51}$Cr | $\leq 1.2 \times 10^{-11}$ | $\leq 2.7 \times 10^{-10}$ |
| $^{51}$V | ML7 | 0.997500 | 1.8 d | 983.55 | $^{51}$V(n,α)$^{48}$Sc | $\leq 1.5 \times 10^{-6}$ | $\leq 1.5 \times 10^{-6}$ |
| $^{58}$Fe | LL1 | 0.002820 | 44 d | 1099.43 | $^{58}$Fe(n,γ)$^{59}$Fe | $\leq 7.0 \times 10^{-11}$ | $\leq 2.5 \times 10^{-8}$ |
| $^{58}$Ni | LL1 | 0.680769 | 71 d | 810.89 | $^{58}$Ni(n,p)$^{58}$Co | $\leq 1.0 \times 10^{-9}$ | $\leq 1.5 \times 10^{-9}$ |
| $^{59}$Co | LL1 | 1.000000 | 5.3 y | 1173 | $^{59}$Co(n,γ)$^{60}$Co | $\leq 5.4 \times 10^{-11}$ | $\leq 5.4 \times 10^{-11}$ |
| $^{63}$Cu | ML3 | 0.691700 | 13 h | 1345.89 | $^{63}$Cu(n,γ)$^{64}$Cu | $\leq 7.0 \times 10^{-9}$ | $\leq 1.0 \times 10^{-8}$ |
| $^{64}$Zn | LL2 | 0.486300 | 244 d | 1116 | $^{64}$Zn(n,γ)$^{65}$Zn | $\leq 8.4 \times 10^{-10}$ | $\leq 1.7 \times 10^{-9}$ |
| $^{71}$Ga | ML2 | 0.398920 | 14 h | 834.03 | $^{71}$Ga(n,γ)$^{72}$Ga | $(9.62 \pm 0.50) \times 10^{-11}$ | $(2.41 \pm 0.12) \times 10^{-10}$ |
| $^{74}$Se | LL1 | 0.008900 | 120 d | 121.17 | $^{74}$Se(n,γ)$^{75}$Se | $\leq 3.8 \times 10^{-12}$ | $\leq 4.2 \times 10^{-10}$ |
| $^{75}$As | ML5 | 1.000000 | 1.1 d | 559.1 | $^{75}$As(n,γ)$^{76}$As | $(4.75 \pm 0.07) \times 10^{-10}$ | $(4.75 \pm 0.07) \times 10^{-10}$ |
| $^{76}$Ge | ML1 | 0.074400 | 11 h | 264.4 | $^{76}$Ge(n,γ)$^{77}$Ge | $(2.98 \pm 0.08) \times 10^{-7}$ | $(4.00 \pm 0.10) \times 10^{-6}$ |
| $^{81}$Br | ML6 | 0.493100 | 1.5 d | 776.5 | $^{81}$Br(n,γ)$^{82}$Br | $\leq 6.1 \times 10^{-12}$ | $\leq 1.2 \times 10^{-11}$ |
| $^{85}$Rb | LL1 | 0.721700 | 19 d | 1076.95 | $^{85}$Rb(n,γ)$^{86}$Rb | $\leq 5.1 \times 10^{-10}$ | $\leq 7.0 \times 10^{-10}$ |
| $^{89}$Y | LL6 | 1.000000 | 106 d | 1836.22 | $^{89}$Y(n,2n)$^{88}$Y | $\leq 1.0 \times 10^{-6}$ | $\leq 1.0 \times 10^{-6}$ |
| $^{93}$Nb | ML2 | 1.000000 | 10 d | 934.49 | $^{93}$Nb(n,2n)$^{92m}$Nb | $\leq 2.3 \times 10^{-7}$ | $\leq 2.3 \times 10^{-7}$ |
| $^{94}$Zr | LL7 | 0.173800 | 64 d | 756.79 | $^{94}$Zr(n,γ)$^{95}$Zr | $\leq 2.4 \times 10^{-9}$ | $\leq 1.4 \times 10^{-8}$ |
| $^{98}$Mo | ML5 | 0.241300 | 2.7 d | 739.5 | $^{98}$Mo(n,γ)$^{99}$Mo | $\leq 1.1 \times 10^{-10}$ | $\leq 4.4 \times 10^{-10}$ |
| $^{103}$Rh | LL7 | 1.000000 | 208 d | 475.15 | $^{103}$Rh(n,2n)$^{102}$Rh | $\leq 7.2 \times 10^{-7}$ | $\leq 7.2 \times 10^{-7}$ |
| $^{109}$Ag | LL2 | 0.481610 | 250 d | 884.71 | $^{109}$Ag(n,γ)$^{110m}$Ag | $\leq 7.1 \times 10^{-11}$ | $\leq 1.5 \times 10^{-10}$ |
| $^{110}$Pd | ML3 | 0.117200 | 7.5 d | 342 | $^{110}$Pd(n,γ,β-)$^{111}$Ag | $\leq 1.8 \times 10^{-9}$ | $\leq 1.5 \times 10^{-8}$ |
| $^{112}$Sn | LL5 | 0.009700 | 115 d | 255.08 | $^{112}$Sn(n,γ)$^{113}$Sn | $\leq 2.0 \times 10^{-9}$ | $\leq 2.1 \times 10^{-7}$ |
| $^{113}$In | LL3 | 0.042900 | 50 d | 190.3 | $^{113}$In(n,γ)$^{114m}$In | $\leq 1.3 \times 10^{-11}$ | $\leq 2.9 \times 10^{-10}$ |
| $^{114}$Cd | ML5 | 0.287300 | 2.2 d | 527.9 | $^{114}$Cd(n,γ)$^{115}$Cd | $\leq 6.4 \times 10^{-11}$ | $\leq 2.2 \times 10^{-10}$ |
| $^{122}$Te | LL3 | 0.008900 | 120 d | 159.02 | $^{122}$Te(n,γ)$^{123m}$Te | $\leq 3.8 \times 10^{-11}$ | $\leq 4.2 \times 10^{-9}$ |
| $^{123}$Sb | LL1 | 0.427900 | 60 d | 1691.15 | $^{123}$Sb(n,γ)$^{124}$Sb | $\leq 2.2 \times 10^{-11}$ | $\leq 5.1 \times 10^{-11}$ |
| $^{127}$I | ML10 | 1.000000 | 12.9 d | 388.47 | $^{127}$I(n,2n)$^{126}$I | $\leq 3.2 \times 10^{-7}$ | $\leq 3.2 \times 10^{-7}$ |
| $^{130}$Ba | ML7 | 0.001060 | 11.5 d | 496.24 | $^{130}$Ba(n,γ)$^{131}$Ba | $\leq 1.2 \times 10^{-11}$ | $\leq 1.2 \times 10^{-8}$ |
| $^{133}$Cs | LL1 | 1.000000 | 2.1 y | 795.99 | $^{133}$Cs(n,γ)$^{134}$Cs | $\leq 2.7 \times 10^{-11}$ | $\leq 2.7 \times 10^{-11}$ |
| $^{139}$La | ML5 | 0.999098 | 1.7 d | 487.02 | $^{139}$La(n,γ)$^{140}$La | $\leq 7.3 \times 10^{-12}$ | $\leq 7.3 \times 10^{-12}$ |
| $^{140}$Ce | LL1 | 0.884500 | 33 d | 145.4 | $^{140}$Ce(n,γ)$^{141}$Ce | $\leq 1.3 \times 10^{-10}$ | $\leq 1.5 \times 10^{-10}$ |
| $^{146}$Nd | LL3 | 0.172000 | 11 d | 531.04 | $^{146}$Nd(n,γ)$^{147}$Nd | $\leq 2.5 \times 10^{-10}$ | $\leq 1.5 \times 10^{-9}$ |
| $^{151}$Eu | LL4 | 0.478100 | 13.5 y | 779.01 | $^{151}$Eu(n,γ)$^{152}$Eu | $\leq 1.4 \times 10^{-11}$ | $\leq 3.0 \times 10^{-11}$ |
| $^{152}$Gd | LL3 | 0.002000 | 240 d | 97.48 | $^{152}$Gd(n,γ)$^{153}$Gd | $\leq 1.2 \times 10^{-12}$ | $\leq 6.1 \times 10^{-10}$ |
| $^{152}$Sm | ML2 | 0.267500 | 1.9 d | 103.18 | $^{152}$Sm(n,γ)$^{153}$Sm | $\leq 3.4 \times 10^{-13}$ | $\leq 1.3 \times 10^{-12}$ |
| $^{159}$Tb | LL4 | 1.000000 | 72 d | 879.4 | $^{159}$Tb(n,γ)$^{160}$Tb | $\leq 1.4 \times 10^{-11}$ | $\leq 1.4 \times 10^{-11}$ |
| $^{164}$Dy | ML4 | 0.281800 | 3.4 d | 1379.38 | $^{164}$Dy → $^{166}$Ho [(b)] | $\leq 1.3 \times 10^{-8}$ | $\leq 4.6 \times 10^{-8}$ |
| $^{165}$Ho | ML2 | 1.000000 | 1.1 d | 1379.47 | $^{165}$Ho(n,γ)$^{166}$Ho | $\leq 7.8 \times 10^{-11}$ | $\leq 7.8 \times 10^{-11}$ |
| $^{169}$Tm | LL3 | 1.000000 | 129 d | 84.25 | $^{169}$Tm(n,γ)$^{170}$Tm [(c)] | $(1.02 \pm 0.08) \times 10^{-10}$ | $(1.02 \pm 0.08) \times 10^{-10}$ |
| $^{170}$Er | ML8 | 0.149300 | 7.5 h | 308.2 | $^{170}$Er(n,γ)$^{171}$Er | $\leq 3.3 \times 10^{-10}$ | $\leq 2.2 \times 10^{-9}$ |
| $^{174}$Yb | ML2 | 0.318300 | 4.2 d | 396.21 | $^{174}$Yb(n,γ)$^{175}$Yb | $\leq 8.8 \times 10^{-12}$ | $\leq 2.8 \times 10^{-11}$ |
| $^{176}$Lu | ML4 | 0.002590 | 6.6 d | 208.36 | $^{176}$Lu(n,γ)$^{177}$Lu | $(3.95 \pm 0.56) \times 10^{-14}$ | $(1.52 \pm 0.21) \times 10^{-11}$ |
| $^{180}$Hf | LL1 | 0.350800 | 42 d | 482.18 | $^{180}$Hf(n,γ)$^{181}$Hf | $\leq 7.4 \times 10^{-12}$ | $\leq 2.1 \times 10^{-11}$ |
| $^{181}$Ta | LL1 | 0.999880 | 114 d | 1189.25 | $^{181}$Ta(n,γ)$^{182}$Ta | $(4.23 \pm 0.18) \times 10^{-12}$ | $(4.23 \pm 0.18) \times 10^{-12}$ |
| $^{184}$Os | LL3 | 0.000200 | 94 d | 646.21 | $^{184}$Os(n,γ)$^{185}$Os | $\leq 9.9 \times 10^{-14}$ | $\leq 5.0 \times 10^{-10}$ |
| $^{186}$W | ML5 | 0.284300 | 1.0 d | 685.76 | $^{186}$W(n,γ)$^{187}$W | $(8.77 \pm 0.99) \times 10^{-12}$ | $(3.09 \pm 0.35) \times 10^{-11}$ |
| $^{187}$Re | ML4 | 0.626000 | 17 h | 155 | $^{187}$Re(n,γ)$^{188}$Re | $\leq 8.7 \times 10^{-12}$ | $\leq 1.4 \times 10^{-11}$ |
| $^{191}$Ir | LL4 | 0.373000 | 74 d | 316.42 | $^{191}$Ir(n,γ)$^{192}$Ir | $(2.00 \pm 0.33) \times 10^{-13}$ | $(5.37 \pm 0.89) \times 10^{-13}$ |
| $^{197}$Au | ML5 | 1.000000 | 2.7 d | 411.7 | $^{197}$Au(n,γ)$^{198}$Au [(d)] | $\leq 4.4 \times 10^{-12}$ | $\leq 4.4 \times 10^{-12}$ |
| $^{198}Pt$ | ML2 | 0.07356 | 3.2 d | 158.41 | $^{198}Pt(n,\gamma)^{199}Pt$ | $\leq 1.9 \times 10^{-11}$ | $\leq 2.6 \times 10^{-10}$ |
| $^{202}Hg$ | LL2 | 0.298600 | 46 d | 279.11 | $^{202}Hg(n,\gamma)^{203}Hg$ | $\leq 3.6 \times 10^{-11}$ | $\leq 1.2 \times 10^{-10}$ |
| $^{203}Tl$ | ML3 | 0.295240 | 12.2 d | 439.31 | $^{203}Tl(n,2n)^{202}Tl$ | $\leq 6.2 \times 10^{-8}$ | $\leq 2.1 \times 10^{-7}$ |
| $^{204}Pb$ | ML9 | 0.014000 | 2.1 d | 279.17 | $^{204}Pb(n,2n)^{203}Pb$ | $\leq 3.9 \times 10^{-8}$ | $\leq 2.8 \times 10^{-6}$ |
| $^{232}Th$ | LL1 | 1.000000 | 27 d | 311.83 | $^{232}Th(n,\gamma,\beta-)^{233}Pa$ | $\leq 1.4 \times 10^{-11}$ | $\leq 1.4 \times 10^{-11}$ |
| $^{238}U$ | ML5 | 0.992745 | 2.4 d | 277.49 | $^{238}U(n,\gamma,\beta-)^{239}Np$ | $\leq 2.2 \times 10^{-11}$ | $\leq 2.2 \times 10^{-11}$ |

a) Possible interference with $^{28}Si(n,\alpha p)^{24}Na$ and $^{76}Ge(n,\gamma)^{77}Ge$
b) The sequential reactions during irradiation are $^{164}Dy(n,\gamma)^{165}Dy$ and $^{165}Dy(n,\gamma,\beta-)^{166}Ho$
c) Possible interference with $^{181}Ta(n,\gamma)^{182}Ta$
d) Possible bias due to suspected problems occurred during the preparation of the standard

**Table 1** Contaminant elements in the Si28-21Pr10.2, part 2 sample. The detection limits, evaluated according to Currie's method, and the standard uncertainties include only the contribution due to counting statistics

## Conclusions

Within the research activity carried out for the $N_A$ determination we tested a differential method based on neutron activation and developed to check the purity of the Avogadro $^{28}Si$ single-crystals. The aim was to reach the minimum detection limits for the maximum number of contaminant elements, without paying close attention to the uncertainty. Significant contaminations, if any, will be quantified later with the required uncertainty. The test sample was a $^{28}Si$ crystal which was grown to verify the enrichment process without taking extreme care of the final purity. The analysis included fifty-nine out of the eighty-nine possible impurity elements.

Since the assumption of a natural abundance of the contaminant elements in the highly enriched sample may be doubtful, the results were given both in terms of mass fraction of the detected isotope, and, in case of natural composition, of the relevant element. The silicon sample was found contaminated by Ge. Moreover, traces of Ga, As, Tm, Lu, Ta, W and Ir were also detected. These results confirmed the capability of the method to identify and to quantify the contaminations. For the remaining elements, the analysis reached detection limits ranging between 1.3 pg/g and 1.5 μg/g.

The future activity will focus on decreasing those detection limits that are still higher than 1 ng/g down to the values reported in literature [7]. This target could be achieved by (i) increasing the irradiation time, (ii) using a detection system with a lower background, and (iii) irradiating the sample with a higher neutron flux. Eventually, a sample of the valuable $^{28}Si$ crystal used to determine the Avogadro constant [5, 13] will be supplied by the PTB to verify its chemical purity.

Moreover, if the amount of the minority isotopes $^{29}Si$ and $^{30}Si$ is reduced at μg/g level, the INAA could be further improved by including those contaminant elements that can be detected via short-lived radionuclides.

## Acknowledgments

This work was jointly funded by the European Metrology Research Programme (EMRP) participating countries within the European Association of National Metrology Institutes (EURAMET) and the European Union. The authors would like to thank the PTB for making available the $^{28}Si$ sample and for providing the isotopic composition and the concentrations of oxygen and carbon, the ECP for the isotopic enrichment, the IChHPS for the conversion to silane, purification and poly-crystal deposition, and the IKZ for the single-crystal growth. Moreover, they are grateful to H. Bettin, NV Abrosimov, AV Gusev and DG Aref'ev for valuable discussions, theoretical, and technical support.

**References**

[1] Mills I M, Mohr P J, Quinn T J, Taylor B N and Williams E R (2006) Redefinition of the kilogram, ampere, kelvin and mole: a proposed approach to implementing CIPM Recommendation 1 (CI-2005). Metrologia 43:227-246

[2] Becker P, De Bièvre P, Fujii K, Glaeser M, Inglis B, Luebbig H and Mana G (2007) Considerations on the future redefinitions of the kilogram, the mole and of other units. Metrologia 44:1-14

[3] Massa E and Mana G (2012) The Avogadro and Planck constants for the redefinition of the kilogram. Rivista del Nuovo Cimento 35:353-388

[4] Becker P, Bettin H, Danzebrink H U, Gläser M, Kuetgens U, Nicolaus A, Schiel D, De Bièvre P, Valkiers S and Taylor P (2003) Determination of the Avogadro constant via the silicon route. Metrologia 40:271-287

[5] Andreas B, Azuma Y, Bartl G, Becker P, Bettin H, Borys M, Busch I, Gray M, Fuchs P, Fujii K, Fujimoto H, Kessler E, Krumrey M, Kuetgens U, Kuramoto N, Mana G, Manson P, Massa E, Mizushima S, Nicolaus A, Picard A, Pramann A, Rienitz O, Schiel D, Valkiers S and Waseda A (2011) Determination of the Avogadro constant by counting the atoms in a $^{28}Si$ Crystal Phys Rev Lett. 106:030801 1-4

[6] Bottger M L, Niese S, Birnstein D and Helbig W (1989) Activation analysis: the most important method for control of purity of semiconductor silicon J Radioanal Nucl Chem 130:417-423

[7] Kim N B, Choi H W, Chun S K, Cho S Y, Woo H J and Park K S (2001) Instrumental neutron activation analysis of silicon wafers using the silicon matrix as the comparator J Radioanal Nucl Chem 248:125-128

[8] Takeuchi T, Nakano Y, Fukuda T, Hirai I, Osawa A and Toyokura N (1997) Determination of trace elements in a silicon single crystal J Radioanal Nucl Chem 216:165-169

[9] Huber A, Bohm G and Pahlke S (1993) Industrial applications of instrumental neutron activation analysis J Radioanal Nucl Chem 169:93-104

[10] Verheijke M L, Jaspers H J J and Hanssen J M G (1989) Neutron Activation Analysis of very pure silicon wafers J Radioanal Nucl Chem 131:197-214

[11] Fujinaga K and Kudo K (1981) Application of instrumental neutron activation analysis in Czochralski silicon crystal growth J Radioanal Nucl Chem 62:195-207

[12] D'Agostino G, Bergamaschi L, Giordani L, Mana G, Massa E and Oddone M (2012) Elemental characterization of the Avogadro silicon crystal WASO 04 by neutron activation analysis. Metrologia 49:696-701

[13] Pramann A, Rienitz O, Schiel D, Schlote J, Güttler B and Valkiers S (2011) Molar mass of silicon highly enriched in 28 Si determined by IDMS. Metrologia 48:S20-S25

[14] Mana G, Rienitz O and Pramann A (2010) Measurement equations for the determination of the Si molar mass by isotope dilution mass spectrometry Metrologia 47:460-463

[15] Zakel S, Wundrack S, Niemann H, Rienitz O and Schiel D (2011) Infrared spectrometric measurement of impurities in highly enriched 'Si28'. Metrologia 48:S14-S19

[16] Currie L A (1968) Limits for qualitative detection and quantitative determination. Anal Chem 40:586-593

[17] Bergamaschi L, D'Agostino G, Giordani L, Mana G and Oddone M (2013) The detection of signals hidden in noise. Metrologia 50:269-276